# Automation Bias in the AI Act

On the Legal Implications of Attempting to De-Bias Human Oversight of AI

Johann Laux*, Hannah Ruschemeier**

* British Academy Postdoctoral Fellow, Oxford Internet Institute, University of Oxford, johann.laux@oii.ox.ac.uk

** Professor for Public Law, Data Protection Law and Law of Digitalisation, University of Hagen, hannah.ruschemeier@fernuni-hagen.de

## Abstract

This paper examines the legal implications of the explicit mentioning of automation bias (AB) in the Artificial Intelligence Act (AIA). The AIA mandates human oversight for high-risk AI systems and requires providers to enable awareness of AB, i.e., the human tendency to over-rely on AI outputs. The paper analyses the embedding of this extra-juridical concept in the AIA, the asymmetric division of responsibility between AI providers and deployers for mitigating AB, and the challenges of legally enforcing this novel awareness requirement. The analysis shows that the AIA's focus on providers does not adequately address design and context as causes of AB, and questions whether the AIA should directly regulate the risk of AB rather than just mandating awareness. As the AIA's approach requires a balance between legal mandates and behavioural science, the paper proposes that harmonised standards should reference the state of research on AB and human-AI interaction, holding both providers and deployers accountable. Ultimately, further empirical research on human-AI interaction will be essential for effective safeguards.

### Acknowledgment

JL and HR contributed equally to this publication.

### Funding Statement

This article is a deliverable of the "The Emerging Laws of Oversight" project, supported by a British Academy Postdoctoral Fellowship (grant no. PF22\220076).

## I. Introduction

Despite its sweeping regulation of Artificial Intelligence (AI) across sectors,[1] the European Union's AI Act (AIA) leaves important normative decisions to AI developers and users.[2] For example, there is no specific guidance on which types or degree of unequal treatment by AI systems are unacceptable.[3] Harmonised standards, intended to aid the implementation of the AIA by providing technical specifications for the design and the development of AI systems are currently being developed.[4] These standards, however, are unlikely to address all of the difficult normative questions which the introduction of AI systems in private and public organisations inevitably raises.[5] Even in combination with harmonised standards, the AIA leaves ample discretion for AI developers and users, and thus relies on local human judgment for its implementation.

In fact, Article 14 AIA prescribes human oversight to mitigate the risks of high-risk AI systems. This reliance on human judgment in oversight is accompanied by a vague regulatory recognition of human fallibility manifested as biased decision-making: Article 14(4b) AIA obliges AI providers to deliver their AI systems to AI deployers in a way that natural persons to whom human oversight is assigned, are enabled '*to remain aware* [emphasis added] of the possible tendency of automatically relying or over-relying on the output produced by an AI system (automation bias)'. In the academic literature, automation bias (AB) describes the psychological phenomenon of human overreliance on automation in decision-making, i.e., the human tendency to overly accept the suggestions of an automated system.[6] AB has the potential to exert significant influence, particularly when legal and institutionalised decision-making structures are predicated on a human decision, but the direction of that decision is in fact provided by a machine.[7] AB has been demonstrated in various applications[8] and is difficult to combat because it is multifactorial: technical, psychological, social, and normative factors can all play a role and interact with each other.[9]

[1] Martin Ebers, 'Truly Risk-Based Regulation of Artificial Intelligence How to Implement the EU's AI Act' [2024] European Journal of Risk Regulation 1.

[2] Johann Laux, Sandra Wachter and Brent Mittelstadt, 'Three Pathways for Standardisation and Ethical Disclosure by Default under the European Union Artificial Intelligence Act' (2024) 53 Computer Law & Security Review 105957; R Mühlhoff and H Ruschemeier, 'Regulating AI with Purpose Limitation for Models' (2024) 1 Journal of AI Law and Regulation 24.

[3] Johann Laux, Sandra Wachter and Brent Mittelstadt, 'Trustworthy Artificial Intelligence and the European Union AI Act: On the Conflation of Trustworthiness and Acceptability of Risk' [2023] Regulation & Governance 1; Hannah Ruschemeier and Jascha Bareis, 'Searching for Harmonised Rules: Understanding the Paradigms, Provisions and Pressing Issues in the Final EU AI Act' (25 June 2024) <https://papers.ssrn.com/abstract=4876206> accessed 3 October 2024.

[4] European Commission, 'Commission Implementing Decision of 22.5.2023 on a Standardisation Request to the European Committee for Standardisation and the European Committee for Electrotechnical Standardisation in Support of Union Policy on Artificial Intelligence' (2023) C(2023) 3215 final.

[5] Laux, Wachter and Mittelstadt (n 2).

[6] Kate Goddard, Abdul Roudsari and Jeremy C Wyatt, 'Automation Bias: A Systematic Review of Frequency, Effect Mediators, and Mitigators' (2012) 19 JAMIA (Journal of the American Medical Informatics Association) 121.

[7] Hannah Ruschemeier and Lukas J Hondrich, 'Automation Bias in Public Administration – an Interdisciplinary Perspective from Law and Psychology' (2024) 41 Government Information Quarterly 101953.

Quite surprisingly for a regulatory law, the AIA explicitly names AB as a potential source of distortion for the decision-making of human oversight agents. Even more surprisingly, AB is the only psychological bias named by the AIA, despite human oversight agents likely being subject to other types of biases, too. AB is one of many cognitive biases such as confirmation bias, availability bias, or algorithm aversion (see further below) that may lead to human error in the oversight of AI systems. Cognitive biases are deviations from standards of rationality that occur systematically in human thinking and reasoning. They may stem from humans' use of heuristics to deal with limitations in data or information processing or a lack of expertise and may have evolved from the design of our brains as neural networks developed for survival in ancestral hunter-gatherer lifestyles.[10] AB in particular is often understood as a lack (or misplacement) of attention.[11] It is neither a new development of digitalisation or specific to AI, nor has it arisen during recent technical developments such as large language models (LLMs). Instead, it is a comparatively old phenomenon, well documented in other settings such as aviation.[12]

The novelty of mentioning a singular cognitive bias in a regulatory law combined with the lack of further explanation in the AIA appears to be a somewhat arbitrary regulatory choice and motivated the research presented in this article. The following sections proceed as follows: Section 2 analyses the normative embedding of the extra-juridical concept of AB within the regulatory logic of the AIA. It will be shown that the AIA's legislative history does not provide an answer as to why AB is the only cognitive bias that is mentioned in the legal text. Section 3 investigates the shared responsibility for human oversight between AI providers and deployers. It argues that the causal factors of AB do not match the legal division of responsibilities and thus creates a tense asymmetry: while the provider is responsible for enabling awareness of AB, the bias manifests itself in the deployer's context. Section 4 shows the difficulties of legally enforcing the obligation of enabling awareness, as proving AB occurred in a given human oversight system is challenging, not least because identifying the unbiased true outcome for many oversight decisions it will be difficult (if not impossible). Finally, Section 5 discusses our findings by asking whether AB is a risk that should itself be regulated to close significant gaps in the protection of EU citizens' health, safety and fundamental rights. To address some of the problems identified, we suggest that future harmonised standards should include a reference to the latest state of scientific research on AB and human-AI interaction. Lastly, we take a broader look and consider the relationship of Article 14 AIA with Article 22 of the General Data Protection Directive (GDPR).

---

8 See e.g. for medicine Goddard, Roudsari and Wyatt (supra note 6).; military Raja Parasuraman and Dietrich H Manzey, 'Complacency and Bias in Human Use of Automation: An Attentional Integration' (2010) 52 Human Factors: The Journal of the Human Factors and Ergonomics Society 381.; human resources Cordula Kupfer and others, 'Check the Box! How to Deal with Automation Bias in AI-Based Personnel Selection' (2023) 14 Frontiers in Psychology 1118723.; national security Michael C Horowitz and L Kahn, 'Bending the Automation Bias Curve: A Study of Human and AI-Based Decision Making in National Security Contexts' [2023] arXiv.org. or public administration Saar Alon-Barkat and Madalina Busuioc, 'Human-AI Interactions in Public Sector Decision-Making: "Automation Bias" and "Selective Adherence" to Algorithmic Advice' [2022] Journal of Public Administration Research and Theory.

9 See the references in: Ruschemeier and Hondrich (n 7).

10 JE (Hans) Korteling and Alexander Toet, 'Cognitive Biases', *Encyclopedia of Behavioral Neuroscience, 2nd edition* (Elsevier 2022) <https://linkinghub.elsevier.com/retrieve/pii/B9780128093245241059> accessed 8 April 2025.

11 Parasuraman and Manzey (supra note 8); Goddard, Roudsari and Wyatt (supra note 6).

12 Parasuraman and Manzey (supra note 8); Amos Tversky and Daniel Kahneman, 'Judgment under Uncertainty: Heuristics and Biases' [1982] Science (New York, N.Y.); Skitka, Linda, Mosier; Kathleen L., and Burdick, Mark, 'Accountability and Automation Bias' [2000] International Journal of Human-computer Studies \/ International Journal of Man-machine Studies.

## II. How Is Automation Bias Referenced in the AIA?

As stated in the introduction, AB is referenced in Article 14 AIA which requires human oversight of high-risk AI systems. This section explains how AB fits into the AIA's regulatory logic of different risk categories (2.1) and analyses how an extra-juridical concept that denotes a cognitive bias will likely be applied within legal contexts (2.2).

### *II.1 Automation Bias and AI*

The regulatory framework established by the AIA adopts a risk-based approach, classifying systems according to the level of risk they present—ranging from unacceptable to low or high—while also accounting for the potential systemic risks posed by general-purpose AI systems. Systematically, AB is addressed within the obligations for high-risk AI systems outlined in Articles 9 et seq. of the third and most extensive chapter of the AIA. These obligations apply to providers and operators of AI systems, and Chapter III also lays out provisions regarding notifying authorities and standardisation norms.

It is reasonable to assume that high-risk systems will have great practical relevance for the European single market. The Commission has assessed that 5-15% of AI systems on the EU market are high-risk systems.[13] Article 6(1) provides a dual qualification for high-risk classification: where an AI system is either a safety component of a product or the product itself falling under the regulatory framework of harmonised product safety legislation listed in Annex I. Alternatively, an AI system may be classified as such if it is determined relevant to the fundamental rights as outlined in Article 6(2) AIA in conjunction with Annex III. The context-based catalogue in Annex III refers to areas relevant to fundamental rights and society, such as education, the awarding of public contracts, asylum, criminal proceedings, and democratic processes. It is notable that the AIA does not consider the fact that systems themselves are particularly conducive to AB due to their design as a criterion for a high-risk classification. Rather, AB is only mentioned in the context of the obligations for high-risk systems in Article 9 et seq. AIA.

The legislative history of the AIA offers minimal insight into the rationale behind the explicit mentioning of AB in the legal text.[14] The reference to AB in Article 14(4)(b) AIA was incorporated into the Commission draft of 21 April 2021. Although 1.1 of the proposal noted the Council called for greater efforts in addressing the risk of bias in their conclusions of 21 October 2020, the discourse remained largely generic. Recitals 40 and 44 of the Commission draft revisited the issue, albeit without providing specific instances of AB. However, a more detailed exposition can be found in the Parliament's opinion:

> [The EP] warns that the capabilities of AI may also pose security risks, as they may lead humans to place such confidence in AI that they trust it more than their own judgement; notes that using a human-in-the-loop approach as a corrective

[13] https://eur-lex.europa.eu/legal-content/EN/TXT/?uri=celex%3A52021SC0084.

[14] For an overview on the legislative history in the context of biases, see : Allessia Chiappetta, 'Navigating the AI Frontier: European Parliamentary Insights on Bias and Regulation, Preceding the AI Act' (2023) 12 Internet Policy Review 1.

> mechanism is not feasible in all cases; notes *that experiments have shown* [emphasis added] that this can elevate the level of autonomy of AI beyond the supporting role for which it was originally designed and means that humans miss opportunities to gain experience and refine their skills and knowledge of AI systems; stresses, therefore, that safety by design and meaningful human oversight based on appropriate training as well as appropriate security and privacy safeguards are required in high-risk AI systems in order to overcome such automation bias.

This explicit recognition of the risk of AB is no longer found in the recitals of the AIA (see recital 73). The explicit reference to experiments, i.e., empirical studies, in the Parliament's opinion is particularly interesting. While the final text of the AIA does not offer much explanation on the psychological research behind AB, we argue below that explicit reference to the current state of psychological research future should be included in future harmonised standards.

## *II.2 Extra-Juridical References in the Law*

From a general regulatory perspective, the AIA's explicit mention of AB raises the question of how extra-juridical concepts such as psychological bias may apply in law. AB is not a legal concept, unlike 'property', 'marriage', or 'association', which protect precisely those legal norms that assign a stock of assets to a person or shape the institutions of marriage or association law.[15] Initially a non-legal factor, AB may take on legally relevant aspects, for example in cases of discrimination.[16] Although many cognitive biases are relevant in law,[17] explicit reference is rare in legal provisions. Whether cognitive biases justify legal intervention remains an open normative question, even where their occurrence can be demonstrated empirically.[18] Thus far, psychological findings on the occurrence of cognitive biases and their mitigation appear most frequently in criminal law and criminology,[19] laws regulating judicial behaviour, and the design of legal institutions such as courts and juries (primarily in, for example, rules of procedure and rules of evidence),[20] as well as laws seeking to 'nudge' public policy by subtly guiding individuals toward making certain (beneficial) choices, for example through changing the default choice or by highlighting the positive outcomes of a certain choice.[21]

---

[15] For privacy as an extra-juridical concept: Christoph Gusy, 'Was Schützt Privatheit? Und Wie Kann Recht Sie Schützen?' (2022) 70 Jahrbuch des öffentlichen Rechts der Gegenwart. Neue Folge (JöR) 415, p. 416.

[16] On biases and anti-discrimination law: Monique Mann and Tobias Matzner, 'Challenging Algorithmic Profiling: The Limits of Data Protection and Anti-Discrimination in Responding to Emergent Discrimination' (2019) 6 Big Data & Society 205395171989580; Sandra Wachter, Brent Mittelstadt and Chris Russell, 'Why Fairness Cannot Be Automated: Bridging the Gap Between EU Non-Discrimination Law and AI' (2021) 41 Computer Law & Security Review 105567; Raphaële Xenidis, 'Tuning EU Equality Law to Algorithmic Discrimination: Three Pathways to Resilience' (2020) 27 Maastricht Journal of European and Comparative Law 736.

[17] See in the context of automated decision-making, for example: Marvin van Bekkum and Frederik Zuiderveen Borgesius, 'Digital Welfare Fraud Detection and the Dutch SyRI Judgment' (2021) 23 European Journal of Social Security 323. We lack the space to reference the vast literature on psychological biases and the law here.

[18] For example: Jennifer Arlen and Stephan W Tontrup, 'Does the Endowment Effect Justify Legal Intervention? The Debiasing Effect of Institutions' [2014] SSRN Electronic Journal <http://www.ssrn.com/abstract=2473758> accessed 11 January 2025.

[19] See, for example: Amina A Memon, Aldert Vrij and Ray Bull, *Psychology and Law: Truthfulness, Accuracy and Credibility* (John Wiley & Sons 2003).

[20] Daniel A Farber and Suzanna Sherry, '18 Building a Better Judiciary' in David E Klein and Gregory Mitchell (eds), *The Psychology of Judicial Decision Making* (1st edn, Oxford University PressNew York 2010) <https://academic.oup.com/book/3404/chapter/144529805> accessed 11 January 2025; Jonathan Baron, 'Heuristics and Biases' in Eyal Zamir and Doron Teichman (eds), Jonathan Baron, *The Oxford Handbook of Behavioral Economics and the Law* (Oxford University Press 2014) <https://academic.oup.com/edited-volume/34475/chapter/292502379> accessed 11 January 2025; Johann Laux, *Public Epistemic Authority: Normative Institutional Design for EU Law* (1st ed, Mohr Siebeck 2022).

To our knowledge, Article 14 AIA establishes the first norm of EU law that explicitly mentions cognitive bias. The legal text itself gives only brief explanation of AB as the 'tendency of automatically relying or over-relying on the output produced by a high-risk AI system (automation bias)' (Article 14(4)(b) AIA). While this definition is legally expedient, it is also scientifically simplified, and does not reflect the decades of psychological research into AB. As the norm requires human oversight agents 'remain aware' (Article 14(4)(b) AI Act) of their susceptibility to AB, proper adherence requires consideration of the current state of empirical AB findings in behavioural psychology. However, psychological research into human-AI interactions is complex, dynamic, and likely subject to a significant degree of scientific uncertainty about the causes of AB in high-risk AI applications. Research on socio-technical systems has demonstrated that the interaction between humans and technology is frequently more complex and divergent than anticipated.[22] Although AB is only one of many psychological phenomena relevant to human-AI interactions, with its inclusion, the AIA answered long-standing calls for greater recognition of behavioural factors in legislation and legal scholarship by considering psychological findings in legal compliance and adjudication.[23]

It is worth comparing the reference to AB to the technical and computational references in the AIA. Through its legal definitions, the AIA embraces technical developments in the field of AI and transfers them into the normative system of law. According to the AIA, a system is considered AI in a legally relevant sense if it meets the criteria defined in Article 3(1) AIA. This does not necessarily have to be congruent with a technical understanding from computer science or other domains. For example, the ECJ also understands anonymisation of data in a legal-normative sense and not in a technical sense.[24] In the case of AB, however, the incorporation of extra-juridical concepts into the legal system is not a matter of computational components, but of a behavioural psychological phenomenon. What both psychological and technical/computational references have in common, however, is that they incorporate the knowledge and methods of expert disciplines into the law: settling issues within these respective domains requires specialised training outside of the law.[25] Risk regulation in EU law has a rich history of incorporating scientific terms into its legal text.[26] As such, AIA does not chart an entirely new path, although it appears to be the first EU regulation to demand consideration of concepts from behavioural research for its implementation.

---

[21] See, for example: Alberto Alemanno and Anne-Lise Sibony (eds), *Nudge and the Law: A European Perspective* (Hart Publishing 2015).

[22] Ben Green, 'The Flaws of Policies Requiring Human Oversight of Government Algorithms' (2022) 45 Computer Law & Security Review 105681; Riikka Koulu, 'Proceduralizing Control and Discretion: Human Oversight in Artificial Intelligence Policy' (2020) 27 Maastricht Journal of European and Comparative Law 720.

[23] Already arguing for a more empirical approach: Peter H Schuck, 'Why Don't Law Professors Do More Empirical Research' (1989) 39 Journal of Legal Education 323. Today, especially scholars of behavioural law and economics make the case for empirical and behavioural research, see further: Christoph Engel, 'Empirical Methods for the Law' (2018) 174 Journal of Institutional and Theoretical Economics (JITE) / Zeitschrift für die gesamte Staatswissenschaft 5.

[24] CJEU, Case C-582/14, *Breyer*, para. 45–49 decided that de-identification must be „reasonably likely" for data to be non-anonymous; rightfully critical on this: Philipp Hacker, 'A Legal Framework for AI Training Data—from First Principles to the Artificial Intelligence Act' (2021) 13 Law, Innovation and Technology 257 (267).

[25] Scott Brewer, 'Scientific Expert Testimony and Intellectual Due Process' (1998) 107 Yale Law Journal 1535. For the legal consideration of technical issues in EU law, see: Anne-Lise Sibony and Eric Barbier De La Serre, 'Expert Evidence before the EC Courts' (2008) 45 Common Market Law Review 941.

## III. Who Is Responsible for De-Biasing Human Oversight?

Article 14 AIA requires providers of high-risk AI systems to design and develop their systems so that they can 'effectively be overseen by natural persons' during their use. Additionally, Article 26(2) AIA requires deployers of AI systems to assign human oversight to natural persons who have the necessary skills and competence.[27] The AIA thus divides the responsibility for enabling effective human oversight between the AI provider, who is responsible for the development and design of the AI system, and the AI deployer, who is responsible for the organisational implementation of human oversight. This raises the question of whether the AIA's division of legal responsibility for human oversight matches the findings in psychological literature on the factors that have been shown to cause AB.

Firstly, if AB occurs in human oversight, then it occurs at the level of the deployer, who in complying with the AIA, assigns oversight functions to natural persons who may be susceptible to AB. However, the obligation to enable awareness of AB is limited to providers in Article 14 AIA. The AIA's focus on the provider is a consequence of the European approach to product safety law and its focus on the product providers. Although discussions in the legislative process included shifting a significant aspect of oversight towards the users and deployers of AI systems,[28] this has not been reflected in the final wording of Article 14 AIA.

Psychological research suggests that AB is not only caused by design features at the system level controlled by the provider, but also by individual, motivational, organisational, and contextual factors controlled by the deployer.[29] This section therefore analyses the division of legal responsibilities for human oversight in the AIA with a view to effectively enabling awareness of (and mitigating) human oversight agents' susceptibility to AB.

### *III.1 AI Providers' Obligation to Enable Awareness of Automation Bias*

The AIA does not provide explicit guidance on how providers should ensure human oversight agents are made aware of their vulnerability to AB. Future harmonised standards may suggest awareness-raising measures, but these have not yet been published. To the best of our knowledge, no legal obligations currently exist in the EU that require making an individual aware of their susceptibility to cognitive bias, rendering a comparative legal assessment within EU law impossible.

---

26 See the contributions in: Michelle Everson (ed), *Uncertain Risks Regulated: Facing the Unknown in National, EU and International Law* (Routledge-Cavendish 2008).

27 Deployer under this provision "means a natural or legal person, public authority, agency or other body using an AI system under its authority except where the AI system is used in the course of a personal non-professional activity".

28 Cornelia Kutterer, 'Regulating Foundation Models in the AI Act: From "High" to "Systemic" Risk' (*MIAI*, 11 January 2024) <https://ai-regulation.com/regulating-foundation-models-in-the-ai-act-from-high-to-systemic-risk/> accessed 19 August 2024.

29 Sarah Sterz and others, 'On the Quest for Effectiveness in Human Oversight: Interdisciplinary Perspectives', *The 2024 ACM Conference on Fairness, Accountability, and Transparency* (ACM 2024) <https://dl.acm.org/doi/10.1145/3630106.3659051> accessed 14 January 2025.

There are, however, efforts to reduce implicit biases such as racial or gender bias in court, especially in the United States of America. Design interventions suggested in this field include reminding human decision-makers of their potential subjectivity, fallibility, and the potential for implicit bias by building informative and educational prompts into the work interface; by slowing the pace of decision-making and reducing task complexity, thereby improving decision-making by allowing deliberative processing of information; designing the work interface to avoid provoking negative emotional states (such as frustration or anger); recording decisions to establish a track-record that may reveal biased decision-making (a single decision will not suffice to establish the occurrence of bias, see further below).[30] These efforts demonstrate that debiasing through enabling awareness is not entirely unprecedented.

However, AB is not a social bias such as race or gender bias. The psychological literature understands AB primarily (albeit not exclusively) as a matter of attention and its (mis)placement.[31] Here, the factors assumed to cause AB include the end user's psychological states (such as tiredness) and psychological traits (such as conscientiousness); the user's accountability for decision outcomes; the user's training; the information presented to the user; the design of the system interface and how it presents that information; environmental constraints; workload and task complexity; as well as the social environment.[32] Assuming that these factors also apply to AB for human oversight, it must be said that the provider cannot control for them all. However, the provider is clearly able to control the design of the AI system and its user interface as well as the content of instructions (including information about potential susceptibility to AB) provided to the deployer. Arguably these are the factors that Article 14(4b) AIA addresses.

There is a considerable risk that providers will interpret their obligation under Article 14(4b) AIA as a simple notification requirement, informing human oversight agents about the existence and potential causes of AB. Section 5 argues that Article 14 AIA should instead be interpreted as demanding effective measures that reduce the likelihood of AB. Which factors are most relevant for each high-risk AI system will have to be decided on a case-by-case analysis. Prescribing specific awareness-enabling measures *a priori* for all potential uses of AI systems risks missing important contextual elements. We will further discuss the viability of standardising the prevention of AB in section 5.

Measures to prevent AB also risk leading to overcorrection and humans overly discounting AI systems' outputs, potentially eroding the advantages of introducing AI systems. If overcorrection becomes systematic, it could introduce a bias against AI outputs among those tasked with oversight (on algorithm aversion, see below, section V.1), counteracting the AIA's goal of removing bias in human oversight. The singular focus on AB in Article 14 AIA may thus over-emphasise the risk of AB to the detriment of addressing other cognitive biases. It thus amounts to a highly selective approach towards correcting an AI's outputs. The focus on AB could be rationally reconstructed as a normative choice to tolerate human error as long as these errors do not shift decision-making power to AI systems. The purpose of human oversight, to mitigate risks posed by AI, may align more closely with scenarios where oversight agents overcorrect rather than undercorrect an AI's output. This may normatively reflect a choice to prioritise human autonomy, even if the price of such autonomy is occasional human error.

---

30 Jerry Kang and others, 'Implicit Bias in the Courtroom' (2012) 59 UCLA Law Review 1125.

31 Parasuraman and Manzey (supra note 8); Goddard, Roudsari and Wyatt (supra note 6). For a recent overview of the psychological literature and its links to human oversight, see: Ruschemeier and Hondrich (n 7).

32 See the references in: Ruschemeier and Hondrich (n 7).

To ensure compliance with Article 14(4b) AIA, AI providers may thus be incentivised to design their AI systems to encourage human oversight agents to overcorrect rather than undercorrect system outputs, for example through notifying the human oversight agent of their doubts in the reliability or correctness of the outcome. However, overcorrecting could be costly for the AI deployer, who will incur the error costs arising from false overcorrection, and for the subjects of the AI-driven decision, whose rights may be violated (assuming that overcorrection can pose risks to safety, health, and fundamental rights). In response, AI providers could go beyond mere notification measures and design their AI systems directly in anticipation of over- or undercorrection through human oversight and produce outcomes with the lowest expected error costs. This would, however, exceed the requirements of the wording of Article 14(4b) AIA.

### *III.2 AI Deployers' Obligation to Enable Awareness of Automation Bias*

The obligation of AI deployers to enable human oversight agents' awareness of their susceptibility to AB presents one normative and one factual question. The normative question asks whether deployers should be obliged in any way to take AB-awareness enabling measures. The answer to this depends on the second, factual question, of whether deployers control the measures which could affect the causal factors of AB. Let us first address the latter question.

As stated in section 3.1, organisational choices such as decision-makers' workload and their work environment have been shown to affect the emergence of AB. Moreover, common debiasing strategies (beyond specifically considering AB) include measures such as educational training on the existence and occurrence of implicit biases, testing individuals' susceptibility to bias before selecting them for their role, and increasing diversity amongst teams.[33] This list of measures is not exhaustive, and some, such as education about the existence of bias, fall within the literal meaning of enabling awareness. Others, such as workload distribution, can lower the likelihood of AB without, however, necessarily raising human oversight agents' awareness of their own potential bias. All these measures lie within the control of deployers (although some may prove to be costly to implement). Factually, deployers thus indeed have measures within their control that can enable awareness of AB and reduce the likelihood of biased oversight decisions.

Normatively, while Article 14 AIA only states the obligations of providers, Article 26 AIA lists the obligations for deployers of high-risk AI systems. Article 26(2) AIA states that deployers 'shall assign human oversight to natural persons who have the necessary competence, training, and authority, as well as the necessary support'. The norm does not explicitly mention AB. What kind of competence, training, and support is necessary should, however, be determined in conjunction with Article 14 AIA. If the European legislator appreciates the risk of AB as so important that it is explicitly mentioned in the AIA's legally binding text on human oversight, then the deployers of high-risk AI systems should likewise be obliged to enable awareness of AB. Through textual and systematic interpretation, it is thus at least possible (and we believe plausible) to assume that deployers have an obligation to enable awareness of AB insofar as they control the causal factors of AB. Whether this obligation should also include measures which reduce the likelihood of AB while not necessarily raising human oversight agents' self-awareness of their susceptibility to AB will be discussed in section 5.

---

[33] Kang and others (supra note 30). See also: Johann Laux, 'Institutionalised Distrust and Human Oversight of Artificial Intelligence: Towards a Democratic Design of AI Governance under the European Union AI Act' (2024) 39 AI & SOCIETY 2853.

## IV. Is the Enabling of Awareness of Automation Bias Legally Enforceable?

The obligation to 'enable awareness' under Article 14(4b) AIA creates further legal uncertainty as regards its enforcement. 'Awareness' is a subjective internal state. In principle, such internal states can be proven by established rules of evidence, such as intent in criminal law and corresponding procedural rules. This raises the question of whether a lack of awareness can be proven as a violation of Article 14(4b) AIA. It would then be sanctionable according to Article 99(4a) AIA in conjunction with Article 16(a) AIA and therefore subject to a substantial fine.

Arguably, the wording of Article 14(4b) AIA only states that deployers must enable human oversight agents' awareness of their susceptibility to AB, in practice, most likely through the design of their AI system and accompanying user instructions. In light of this, and the role of the deployer's organisational choices in lessening susceptibility to AB discussed in the previous section, it could be argued that the provider is not responsible for whether human oversight agents are actually aware how prone they are to AB. However, we reject this view, given Article 14 AIA explicitly requires human oversight to be effective, thereby establishing effectiveness as the standard for legal assessment. A failure to assess whether measures intended to raise awareness of bias actually achieved this goal renders the requirement for effective human oversight meaningless. Moreover, presuming deployers' responsibilities include effectively making their oversight agents aware of AB, the question of how exactly how a lack of awareness could be proven arises again in the enforcement of Article 26 AIA.

If providers (and deployers) bear the responsibility for effectively enabling awareness of AB, then the occurrence of AB could serve as evidence of failure to meet this obligation. Ultimately, the aim of fostering awareness of susceptibility to bias is reducing the occurrence of bias.

### *IV.1 How to Prove Automation Bias Has Occurred?*

What then are the prospects for proving *ex post* that AB occurred in a particular human oversight process? A legal procedure will typically focus on a specific case of oversight failure that may involve a single human decision or multiple human decisions. However, a single decision will rarely be sufficient to demonstrate that AB influenced the oversight outcome. Establishing the occurrence of psychological bias usually requires multiple data points. Moreover, psychological research on AB is commonly experimental.[34] While a particular human oversight process may be experimentally tested for its susceptibility to AB, such *ex post* reenactments are likely prohibitively costly and time-consuming. In the future, standardised tests developed for AB could lower the costs of assessment. In psychological research, there are assessments for unconscious biases such as the Implicit Association Test (IAT) to determine whether unconscious biases cause discriminatory decision outcomes.[35] However, the psychological literature still debates the predictive value of the IAT[36] which potentially lowers the probative value of unconscious bias tests in legal procedures. Moreover, as mentioned in section 3.1 above, the psychological literature understands AB primarily as a matter of attention, not of discriminatory association. Whether standardised tests for AB will become available and be of value for legal procedures in the future must thus remain an open question.

[34] Skitka, Linda, Mosier; Kathleen L., and Burdick, Mark (n 12); Max Schemmer and others, 'On the Influence of Explainable AI on Automation Bias' (1 April 2022) <https://ui.adsabs.harvard.edu/abs/2022arXiv220408859S> accessed 29 January 2025.

[35] Anthony G Greenwald, Debbie E McGhee and Jordan LK Schwartz, 'Measuring Individual Differences in Implicit Cognition: The Implicit Association Test' (1998) 74 Journal of Personality and Social Psychology 1464; Jeffrey J Rachlinski and Sheri L Johnson, 'Does Unconscious Racial Bias Affect Trial Judges' (2009) 84 Notre Dame Law Review 1195.

These obstacles to empirical testing suggest that expert testimony based on a review of the oversight conditions may offer the more practical route. Effective risk governance measures such as technical documentation (Article 11 AIA), record-keeping (Article 12 AIA), and transparency (Article 13 AIA) would thus be crucial to document oversight decisions for expert review. For example, experts could consider the information on AB provided to human oversight agents or check time stamps for oversight decisions for signs of attentional deficits.

Evidence provided by expert of behavioural psychology may prove crucial should the AB awareness requirement become relevant in a legal proceeding. As the Court of Justice of the European Union (CJEU) understands its institutional role as the prime interpreter of EU law,[37] it is reluctant to engage in its own fact finding and relies heavily on submissions from the parties.[38] In past proceedings featuring a significant degree of scientific uncertainty (as it may arguable be the case with research on AB in human-AI interaction), the CJEU paid attention to the credentials of the parties' experts, their scientific domains, and whether they work for public or private institutions.[39] Expert testimony is therefore the most probable way in which the occurrence of AB would be proven in legal proceedings.

## *IV.2 Do We Know the Unbiased True Outcome?*

Another complication arises in judging whether decisions are biased, as this presupposes knowing the unbiased true outcome (i.e., the ground truth). Article 11 AIA suggests that AI providers inform deployers about how their AI systems should perform. The technical documentation for AI systems demanded by the AIA includes information on 'degrees of accuracy for specific persons or groups of persons', the 'overall expected level of accuracy' and 'foreseeable unintended outcomes and sources of risks to health and safety, fundamental rights and discrimination' (No. 3 Annex IV of the AIA).[40] Deployers' human oversight agents thus seem to be responsible for monitoring the accuracy of AI systems measured against the accuracy expected by the providers. However, according to Article 14 AIA, the scope of human oversight clearly goes beyond errors of accuracy and includes the protection of fundamental rights.[41] Let us consider an example of non-discrimination and fairness in hiring.

---

36 Jens Agerström and Dan-Olof Rooth, 'The Role of Automatic Obesity Stereotypes in Real Hiring Discrimination.' (2011) 96 Journal of Applied Psychology 790; Frederick L Oswald and others, 'Predicting Ethnic and Racial Discrimination: A Meta-Analysis of IAT Criterion Studies.' (2013) 105 Journal of Personality and Social Psychology 171.

37 Ditlev Tamm, 'The History of the Court of Justice of the European Union Since Its Origin' in Court of Justice of the European Union (ed), *The Court of Justice and the Construction of Europe: Analyses and Perspectives on Sixty Years of Case-law - La Cour de Justice et la Construction de l'Europe: Analyses et Perspectives de Soixante Ans de Jurisprudence* (T M C Asser Press 2013) 12 <http://link.springer.com/10.1007/978-90-6704-897-2_2> accessed 17 January 2025; Takis Tridimas, *The General Principles of EU Law* (2. ed., 1. publ. in paperback, Oxford Univ Press 2007) 216.

38 Tridimas (supra note 35) 216.

39 Laux (n 20) 240–264. The EU AI Office could play an important role in monitoring the field of psychological research on human-AI interaction, see Hadrien Pouget and Johann Laux, 'A Letter to the EU's Future AI Office' (3 October 2023) <https://carnegieendowment.org/2023/10/03/letter-to-eu-s-future-ai-office-pub-90683> accessed 25 January 2024.

40 AIA Annex IV 3.

Imagine, for example, that the provider of an AI system for selecting job interview candidates provides technical information on minimum selection rates for minority groups as their (self-imposed) ground-truth for detecting unfairness. The provider sets the selection rate for minority groups at 80% of that of the group with the highest selection rate.[42] The provider further states that their AI system expectedly performs within the 80% threshold for all groups. As *Langer et al.* (who are drawing on the same example) show, a human oversight agent who detects a selection rate of female candidates of 0.30 and of male candidates of 0.40 would thus have numerical evidence of unfairness in the AI's outputs (as the minority rate is at 75% of the best-performing group)–but whether they will judge this output to actually reflect unfairness will often be left to the agents' discretion or that of their supervisor.[43] If there are ten male and ten female initial candidates altogether and three female and four male candidates are invited for the job interview, then this would signal unfairness according to the ground-truth for fairness set by the provider. However, as *Langer et al.* show, the deployer may still judge the process to have been fair after all: they may not be responding to the numerical evidence as sufficient evidence for unfairness in the selection process.[44]

Would the deployers' human oversight agent be bound by the fairness interpretation of the provider? To protect the fundamental rights of the job candidates as Article 14 AIA demands, the human oversight agent would not have to adhere to the providers' statistical measures of fairness but instead consider the requirements of EU non-discrimination law.[45] Thus far, EU non-discrimination law has relied less on statistical evidence to prove *prima facie* discrimination, instead defining the requirements for legally classifying the treatment of a group as discriminatory by context and judicial intuition.[46]

This demonstrates how the ground-truth for fairness and non-discrimination can vary between providers and deployers as well as the EU legislature and judiciary. This potential drift of the ground truth reveals the instability of our understanding of the unbiased true outcome for an important subset of tasks of human oversight agents. In turn, this significantly complicates the determination of whether a human oversight outcome has been unduly influenced by AB. Moreover, in legal proceedings (a high-risk domain for AI systems according to the AIA), there is no procedure-independent ground-truth available at all: the only criterion for assessing the outcome of a legal procedure is another legal procedure.[47] There are no criteria external to the legal procedure that can establish the correctness of a judicial decision.[48]

---

[41] Markus Langer, Kevin Baum and Nadine Schlicker, 'Effective Human Oversight of AI-Based Systems: A Signal Detection Perspective on the Detection of Inaccurate and Unfair Outputs' (2024) 35 Minds and Machines 1.

[42] This 80% rule is taken from the example in: Leo Alexander, 'Making Decisions about Adverse Impact: The Influence of Individual and Situational Differences' (Rice University 2022) <https://hdl.handle.net/1911/114147> accessed 15 January 2025. Cited after: Langer, Baum and Schlicker (supra note 39).

[43] Langer, Baum and Schlicker (supra note 39) 12–13.

[44] ibid 13.

[45] Wachter, Mittelstadt and Russell (supra note 14).

[46] ibid.

[47] On this problem and jury decisions, see: Jon Elster, *Securities against Misrule: Juries, Assemblies, Elections* (Cambridge university press 2013). For an application to the EU legal system, see: Laux (n 20). For the context of human oversight, see: Langer, Baum and Schlicker (supra note 39) 12.

[48] John Anthony Jolowicz, *On Civil Procedure* (Cambridge university press 2000) 86.

In sum, if providers and deployers are obliged to effectively raise awareness of AB and thus reduce the likelihood of AB, then legal enforcement of this obligation faces considerable challenges. It appears that for the time being, expert testimony is the only viable method for proving AB has occurred or that a given human oversight scheme is prone to AB.

## V. Discussion

Our analysis of the AIA revealed that the obligation to enable awareness of AB can arguably be assigned to both the provider and the deployer of high-risk AI systems. The AIA, however, only explicitly mentions AB in stating the obligations of providers. It also remains unclear what kind of measures providers (and deployers) should take to address AB. We have also shown how further empirical challenges await the enforcement of the enabling of awareness and require deference to experts in psychology. This section discusses our findings with a view on strengthening the protection of EU citizens' health, safety, and fundamental rights and by including the broader regulatory landscape, namely the GDPR.

### *V.1 Is Automation Bias a Risk that Should Itself Be Regulated?*

If the explicit mention of AB in Article 14 AIA indicates the importance the European legislator attaches to debiasing human oversight, should the AIA then directly demand a reduction in the occurrence of AB rather than merely enabling awareness of AB? One could read the weak formulation in Article 14 AIA as a reflection of the scientific status quo: the psychological literature on AB in human-AI interaction is still emerging, with commensurate scientific uncertainty about the causes of (and remedies for) AB in high-risk AI use cases. Moreover, legal scholars do not agree on how the law can be utilised to reduce cognitive biases. Some have argued that 'demonstrated cognitive biases have grown like weeds in a vacant lot'.[49] How the multitude of biases that potentially apply to a given context interact with one another has not been well understood,[50] and AB is not the only cognitive bias relevant to human-AI interaction. Algorithm aversion, for example, denotes humans' tendency to trust an algorithm's opinion less than that of a human even when the algorithm is shown to be more accurate.[51] How do AB and algorithm aversion interact? Do they cancel each other out? The AIA circumnavigates the problem of detailing an under-studied psychological phenomenon at the cost of legal uncertainty about how to enforce the enabling of AB awareness (see section 4 above). Similarly, not all measures which could be taken to reduce cognitive bias necessarily raise awareness of a susceptibility to bias. The AIA thus unnecessarily limits the scope of debiasing measures for AB to awareness-raising measures. We therefore suggest that the AIA's harmonised standards should reference the latest state of scientific research on AB, as we will explain further in the next sub-section. The remainder of this section highlights a legal gap that will continue to exist if the effects of AB are not considered further.

---

[49] William N Eskridge and John Ferejohn, 'Structuring Lawmaking to Reduce Cognitive Bias: A Critical View' (2002) 87 Cornell Law Review 616.

[50] ibid. For more on biases in legal institutional design: Laux (n 20) 267–271.

[51] Berkeley J Dietvorst, Joseph P Simmons and Cade Massey, 'Algorithm Aversion: People Erroneously Avoid Algorithms after Seeing Them Err.' (2015) 144 Journal of Experimental Psychology: General 114; Ryan P Kennedy, Philip D Waggoner and Matthew M Ward, 'Trust in Public Policy Algorithms' (2022) 84 The Journal of Politics 1132. For the context of the AIA, see further: Laux, Wachter and Mittelstadt (n 3).

Article 6 AIA defines the classification methodology for high-risk systems. A system can be classified as high risk either because it is a product or a safety component under the NLF (Article 6(1) AIA) or because it falls under the use-cases of Annex III (Article 6(2) AIA). However, Article 6(3) AIA introduces a substantive exception: systems usually referred to in Annex III (education, credit scoring, public services etc.) are not considered high-risk where there is no significant risk of harm to health, safety, or fundamental rights, as determined by a provider's self-assessment before the system is launched on the market. This is supposed to be the case when one of the conditions of Article 6(3)(a)-(d) AIA apply. For some of these conditions however, AB may pose a significant problem. For example, Article 6(3)(c) AIA excludes an AI system from the high-risk category if the system 'is not meant to replace or influence the previously completed human assessment, without proper human review.' However, even if the system is meant to only 'support' a human decision, AB may still cause humans to unduly over-rely on the AI's suggestion. The concern expressed in Article 14 AIA about AB is not reflected in these criteria, which extensively broadens the scope of Article 6(3) AIA.[52] It is also unclear how to prove a human decision was not influenced by the AI system. The fact that the exception in Article 6(3) AIA relies on self-assessment by the providers without *ex ante* review by supervisory authorities creates a notable protection gap: the criteria do not define standards for the role and responsibilities of the human making the decision, nor do they provide standards for verifying compliance. Article 14(4b) AIA does not address these concerns, as its application is contingent on the AI system being deemed high-risk, a premise that the exception under Article 6(3) AIA specifically negates. One way to address this issue would be to require an explicit reference to the obligations under Article 14(4b) AIA in the self-assessment under Article 6(3). Providers would then only be allowed to deviate from the risk classification if they have checked the safeguards against the AIA. Whether this adequately addresses the risks is an open empirical question, but such an obligation would be legally and logically consistent.

## *V.2 Harmonised Standards Should Reference the State of Research*

The legal uncertainty about how to reduce AB in human oversight is inherently intertwined with the challenging regulatory structure of the AIA. The AIA merges two fundamentally different regulatory approaches: product safety law and the protection of fundamental rights. The fundamental mismatch between the European product safety regime and fundamental rights protection raises normative concerns. The latter relies on proportionality assessments to provide the greatest possible protection from risk for an individual's rights. Product safety law, on the other hand, assumes harm can be measured quantitively[53] and relies on a formalised logic of evaluation against a baseline of acceptable risk.[54] References to AB make these already complex relationships even more complicated by explicitly referring to extra-juridical factors while also requiring providers to enable 'proportionate' (Article 14(4) AIA) awareness of AB. Harmonised standards provide an alternative forum for addressing AB under the AIA, offering more room for extra-juridical factors to be included in the standardised text.

[52] For a critique of this loophole: Hannah Ruschemeier, 'Art. 6 KI-VO' in Mario Martini and Christiane Wendehorst (eds), *Kommentar zur KI-VO* (Beck 2024); Sandra Wachter, 'Limitations and Loopholes in the EU AI Act and AI Liability Directives: What This Means for the European Union, the United States, and Beyond' (2024) 26 Yale Journal of Law & Technology 671.

[53] On the possibility and limitations of including quantitative scientific information in proportionality analyses, see: Laux (n 20) 225–240.

[54] Marco Almada and Nicolas Petit, 'The EU AI Act: A Medley of Product Safety and Fundamental Rights?' (European University Institute 2023) Working Paper <https://cadmus.eui.eu/handle/1814/75982> accessed 1 February 2024; Ruschemeier and Bareis (n 3).

The legal definitions in the AIA embrace technical developments in the field of AI and transfer them into the normative system of law (see section 2.2). In the introduction, we also mentioned that standards will aid the implementation of the AIA by providing further technical specifications. This logic suggests harmonised standards under the AIA that address human oversight should at least include a reference to the 'state of research' on AB. Such a state-of-research clause would establish a qualitative benchmark for addressing bias in human oversight that is itself dynamic and open to development through empirical research. Ideally, the clause would deviate from the narrow focus on AB in Article 14 AIA and reference the state of research on human-AI interaction, thus including all relevant cognitive biases. The concrete mentioning of AB at the level of the AIA seems overly concrete for a law that aims to regulate AI systems across sectors. AB may prove to be a minor issue for human oversight in some contexts in which other cognitive biases may exert a stronger influence on human decision-makers. It would have been a sounder regulatory approach to counteract all cognitive bias in human oversight, allowing for context-sensitive implementation and concretisation.[55] State-of-research or state-of-technology clauses are common in regulatory law and could directly be included in the wording of Article 14 AIA.[56] Since the AIA is already in force, however, this article's reliance on harmonised standards reflects a pragmatic consideration: an amendment of the AIA's text is rather unlikely to happen anytime soon, whereas standards are still being drafted. Technical standardisation itself is, of course, far from perfect. It may be captured by industry interests and again raises the question of whether standards should entail detailed rules or flexible principles.[57] The proposed state-of-research clause would at least open the legal implementation of the AIA to incorporating findings from social science.

Moreover, as mentioned above, the current state of research on AB would likely reveal a significant degree of scientific uncertainty about causes, remedies, and interactions with other cognitive biases. The anticipated effect of uncertainty is tighter judicial scrutiny of the credentials and provenance of the social scientists involved in AB research, with stronger weight attached to findings from public (or at least industry-independent) research institutions (see above, section IV.1).

The state-of-research clause would thus require providers (and, as we argue, ideally also deployers) to continuously update and verify their knowledge on the causes of and remedies to AB. For example, the 'gold standard' for addressing AB could lie in empirically testing the potential for AB in each human oversight scheme, although the European legislator and European standardisation bodies will likely have little appetite for prescribing such a costly measure for all high-risk AI systems. Alternatively, standard(s) may include a checklist on causes of and remedies for AB based on the current state of research, as verified by experts from psychology.

---

[55] On guiding human behaviour in working with AI systems through concrete rules or vague standards, see also: Johann Laux, Fabian Stephany and Alice Liefgreen, 'Improving Task Instructions for Data Annotators: How Clear Rules and Higher Pay Increase Performance in Data Annotation in the AI Economy' (arXiv, 2023) <https://arxiv.org/abs/2312.14565> accessed 18 June 2025.

[56] On such clauses in German law and EU law, cf. Mark Seibel, 'Abgrenzung Der „allgemein Anerkannten Regeln Der Technik" Vom „Stand Der Technik"' [2013] Neue Juristische Wochenschrift 3000.

[57] Laux, Wachter and Mittelstadt (n 2).

## *V.3 Relationship with the GDPR*

The divided responsibility between Article 14 and 26 AIA also raises questions about the relationship with the GDPR. The AIA applies in principle alongside the GDPR and does not affect its scope of application (Article 2(7) AIA). Article 22 GDPR, the prohibition of fully automated decisions when processing personal data, has so far overlooked the issue of AB. The issue is twofold: exact requirements for the level of human involvement with the machine decision proposal remain unclear,[58] and externally proving a decision was actually made by a human is almost impossible. Article 22 GDPR additionally provides exceptions for situations where the data subject is not to be protected from fully automated decisions. According to paragraph 2(a), the prohibition does not apply if the decision is necessary for entering into or performing a contract, (b) the decision is authorised by Union or national law and provides for adequate safeguards for the rights and legitimate interests of the data subject, or (c) the decision is based on consent. However, these exceptions apply to situations in which the interests of the data subjects are otherwise protected, either through voluntary entry into a contract, a decision by the legislator, or through their own consent.

Although the regulations are fundamentally different, the requirement for human oversight in the AIA pursues a similar goal to that of Article 22 GDPR: affected parties should not be subject to automated decisions without the possibility of human intervention. The joint EDPB-EDPS opinion on the AIA emphasises the intersection of the provisions and states that qualified human oversight could ensure that the Article 22 GDPR requirement—'the right not to be subject to a decision based solely on automated processing'—is respected.[59]

Nevertheless, and in light of the ECJ's *SCHUFA* decision, we argue that compliance with the (unclear) requirements of Articles 14(4b) and 26(2) AIA does not rule out an automatic decision under Article 22 GDPR. The SCHUFA system itself is unlikely to fall under the AIA, as it is not an AI system under Article 3(1) AIA. However, according to the ECJ, the criterion for an automated decision under Article 22 GDPR is whether the subsequent final decision (in this specific case, the granting of the loan) was decisively based on the preceding automated decision, even if the responsible person has formal and substantive decision-making powers.[60] In issuing this criterion, the ECJ took a comprehensive view of the decision-making situation focusing on more than the formal responsibility of the person involved, to include (for this case) the entire process of granting a loan. The AIA, on the other hand, only addresses the design decisions of the provider or the deployer: Consequently, a decision may still constitute an automated decision under Article 22 GDPR even if the provider and deployer fulfil the requirements of Articles 14(4b), 26(2) AIA, and responsible persons are made aware of AB.

---

[58] See discussion in: Marco Almada, 'Human Intervention in Automated Decision-Making: Toward the Construction of Contestable Systems', *Proceedings of the Seventeenth International Conference on Artificial Intelligence and Law* (Association for Computing Machinery 2019) <https://doi.org/10.1145/3322640.3326699> accessed 13 January 2025; Stephan Dreyer, Wolfgang Schulz, and Bertelsmann Stiftung, 'The General Data Protection Regulation and Automated Decision-Making: Will It Deliver?: Potentials and Limitations in Ensuring the Rights and Freedoms of Individuals, Groups and Society as a Whole' [2019] Discussion paper Ethics of Algorithms 18 <https://www.bertelsmann-stiftung.de/doi/10.11586/2018018> accessed 13 January 2025; Gianclaudio Malgieri and Giovanni Comandé, 'Why a Right to Legibility of Automated Decision-Making Exists in the General Data Protection Regulation' (2017) 7 International Data Privacy Law 243, 253.

[59] EDPB-EDPS Joint Opinion 5/2021 on the proposal for a Regulation of the European Parliament and of the Council laying down harmonised rules on artificial intelligence (Artificial Intelligence Act), p.6.

[60] ECJ C-634/21, ECLI:EU:C:2023:957.

The responsibilities of the AIA are not the same as those of the GDPR. Responsibility for the processing of personal data by the deployer at the stage of actual application of an AI system is relatively clear: the deployer is the controller under Article 4(7) GDPR. However, the provider also being subject to specific obligations under Article 14(4b) AIA to ensure human oversight and measures against AB, suggests a joint controllership between provider and deployer under Article 26 GDPR. This would also be in line with the complementary obligations under Article 14(4b) and Article 26(2) AIA. In practice however, those who only provide the AI system do not have any influence on the specific data processing carried out by those deploying it.[61] In absence of a specific contractual arrangement, there is thus no joint controllership under Article 26 GDPR. Provider and deployer remain two different data processors. Therefore, it remains unclear how a precise distinction of responsibility could be made either in terms of a breach of Article 14(4b) AIA or Article 26(2) AIA.

## VI. Conclusion

Our analysis of AB in the AIA raises critical questions about the justiciability of the requirement to enable human oversight agents' self-awareness of their susceptibility to AB. Moreover, it remains uncertain whether the legislator considers mere awareness of AB to be an adequate remedy or whether additional measures are required to mitigate its effects. If awareness alone proves to be insufficient, this article argued that the legal obligation under Article 14 AIA for human oversight to be effective would require AI providers (and deployers) to adopt proportionate and evidence-based debiasing interventions aligned with the most recent and state of the art research on AB. Such an obligation would go beyond fostering awareness, demanding active engagement with strategies to counteract bias at the design and implementation stages of AI systems. This would also require aligning the currently asymmetric division of responsibilities between AI providers and deployers for AB awareness with their actual influence on the causal factors of AB. Furthermore, it remains unclear why AB is the only cognitive bias explicitly mentioned in the legal text of the AIA. This focus on AB alone may suggest a prioritisation of human autonomy: the AIA deems human oversight agents erroneously rejecting an AI's recommendation as less problematic than human oversight agents erroneously following an AI's recommendation. However, if the European legislator is serious about debiasing human oversight, it should standardise a reference to the latest state of research in behavioural research on human-AI interaction, including all relevant cognitive biases and holding both providers and deployers responsible.

In conclusion, addressing the unconscious nature of AB through legal obligations requires a nuanced approach, balancing legislative intent, the latest empirical insights, and the practical realities of AI deployment. For these measures to be effective and justiciable, the interplay between legal mandates and cognitive science must be carefully navigated, ensuring awareness translates into meaningful, enforceable safeguards. For now, the implementation of Article 14(4b) AIA must navigate a considerable degree of both scientific and normative uncertainty.

---

[61] Although the ECJ has ruled that the provision of an infrastructure for data processing is sufficient, in the specific case the application used by the authority was personalised and developed for a specific purpose; EC, C-683/21; J ECLI:EU:C:2023:949; par. 27 et. seq.; In the ECJ's IAB Europe case (C-604/22, ECLI:EU:C:2024:214, par. 67 et. seq.), the contribution to data processing was that the IAB association provided a binding regulatory framework for its members, specifying how those members must store and process personal data. These considerations cannot be seamlessly applied to the use of AI systems.

## Competing Interests Declaration

The author(s) declare(s) to not have any competing interests.